**Assessing the cognitive consequences of the object-oriented approach: a survey of empirical research on object-oriented design by individuals and teams**

Françoise Détienne

Ergonomic Psychology Project

INRIA

Domaine de Voluceau,

Rocquencourt, BP 105

78153, Le Chesnay Cedex

France

Francoise.Detienne@inria.fr

**Abstract**

This paper presents a state-of-art review of empirical research on object-oriented (OO) design. Many claims about the cognitive benefits of the OO paradigm have been made by its advocates. These claims concern the ease of designing and reusing software at the individual level as well as the benefits of this paradigm at the team level. Since these claims are cognitive in nature, its seems important to assess them empirically. After a brief presentation of the main concepts of the OO paradigm, the claims about the superiority of OO design are outlined.

The core of this paper consists of a review of empirical studies of OOD. We first discuss results concerning OOD by individuals. On the basis of empirical work, we (1) analyse the design activity of novice OO designers, (2) compare OO design with procedural design and, (3) discuss a typology of problems relevant for the OO approach. Then we assess the claims about naturalness and ease of OO design.

The next part discusses results on OO software reuse. On the basis of empirical work, we (1) compare reuse in the OO versus procedural paradigm, (2) discuss the potential for OO software reuse and (3) analyse reuse activity in the OO paradigm. Then we assess claims on reusability.

The final part reviews empirical work on OO design by teams. We present results on communication, coordination, knowledge dissemination and interactions with clients. Then we assess claims about OOD at the software design team level.

In a general conclusion, we discuss the limitations of these studies and give some directions for future research.

**1.     Introduction**

This paper presents a state-of-art review of empirical research on object-oriented (OO) design. OO technology has been accepted by both academia and industry in a very short time. Ten years ago, the few OO languages available were seen as interesting ideas. Since then, new languages have been designed and existing languages have been extended to become object-oriented. Many claims about the cognitive benefits of the OO paradigm have been made by its advocates. These claims concern the ease of designing and reusing software at the individual level as well as the benefits of this paradigm at the team level. Given the rapid acceptance of this new technology, it is imperative to understand the process and to evaluate the benefits of OO design. Since these claims are cognitive in nature, its seems important to assess them empirically (Détienne & Rist, 1995).

Since 1990, much empirical work has been conducted on OO design both at the individual level and at the software design team level. These studies are field studies and laboratory studies. They are mostly focused on upstream activities, e.g., design, analysis, and communication, rather than on downstream activities, e.g., coding. This paper will present an overview of the results of these studies. After a brief presentation of the main concepts of the OO paradigm, the claims about the superiority of OO design are outlined.

The core of this paper consists of a review of empirical studies of OOD. In section 4, we present results on OO design by individuals. On the basis of empirical work, we (1) analyse the design activity of novice OO designers, (2) compare OO design with procedural design and, (3) discuss a typology of problems relevant for the OO approach. In a final section we assess the claims on naturalness and ease of use of OO design.

Section 5 discusses results on OO software reuse. On the basis of empirical work, we (1) compare reuse in the OO versus procedural paradigm, (2) discuss the potential for OO software reuse and (3) analyse reuse activity in the OO paradigm. In a final section, we assess claims on reusability.

Section 6 reviews empirical work on OO design of teams. We present results on communication, coordination, knowledge dissemination and interactions with clients. In a final section, we assess claims about OOD at the software design team level.

In a general conclusion, we discuss the limitations of these studies and give some directions for future research.

**2.      Concepts of the OO paradigm**

The main concepts of the OO paradigm are the concepts of class, inheritance, and encapsulation. A key difference between the object-oriented paradigm and the procedural paradigm is that in the procedural paradigm data and functions are separated, whereas in the object-oriented paradigm they are integrated. Objects are program entities which integrate a structure defined by a type and functionalities. The concept of <u>class</u> integrates both the structure and behaviour of objects. Objects are instances of classes. Attributes and methods are defined for the entire class. A class is defined as a structure (a type) and a set of methods. A method is a function attached to a class that describes a part of the behaviour of the objects which are instances of this class.

Using <u>inheritance</u>, a hierarchy of classes can be constructed in which the most general attributes and methods are specified in the higher level classes and are inherited by more specialised classes,

In OO programming, <u>abstraction</u> is obtained by the means of encapsulation, polymorphism and late binding. <u>Encapsulation</u> means an object owns its data and methods. The data and methods are private and may be accessed and used by other objects only if the other objects send an appropriate message to the owner. The using object may send the same message to multiple objects which will act on it differently according to their own interpretations. This is the property of <u>polymorphism,</u> which aids abstraction by allowing messages to remain abstract. It is only during execution that the system decides which method will be executed according to the object with which the method is called. This is the property of <u>late binding</u>.

 

According to Rist (Rist, 1994; Rist, 1996; Rist & Terwilliger, 1995), plans and objects are orthogonal in OO systems. This reflects the real world, where a plan can use many objects (the plan for a cake uses flour, eggs, water, and so on), and an object can be used in many plans (an egg can be used to make a cake, omelette, soufflé and so on). A plan is a set of actions that, when placed in a correct order, achieve some desired goal. In an OO system, the actions in a plan are encapsulated in a set of routines, and the routines are divided among a set of classes and connected by control flow.

### 3.  Claims about the benefits of the OO paradigm

Advocates of OO design have made strong claims about the naturalness, ease of use, and power of this design approach (see for example: Meyer, 1988; Rosson and Alpert, 1990). Claims are also made about reusability, and benefits at the software design team level.

### 3.1  Claims about naturalness and ease of OO design

It is claimed that mapping between the problem domain and the programming domain should be more straightforward in the OO paradigm than in the procedural paradigm. The theoretical argument in support of OO design is that objects are clear and visible entities in the problem domain. They are represented as explicit entities in the solution domain, and thus the mapping between the problem and solution domains is simple and clear. The domain objects are identified and used to structure the software system. As a result, the OO paradigm should entail a greater focus on the problem. It should facilitate object-based problem decomposition as well as object-based design solutions.

Concerning problem understanding, it is assumed that the identification of objects (or classes, as objects are the instances of classes) should be easy because objects form natural representations of problem entities. According to Meyer (1988), the world can be naturally structured in terms of objects; thus, it seems particularly relevant to organise a model of design around a software representation of these objects.

With an OOP language, decomposing the problem into a solution consists of identifying the relations between objects and the associations between their structures and the functionalities. It is assumed that decomposing a problem into objects is natural. This activity is assumed to be driven much more by a designer's knowledge about world structure than by knowledge about the design process or about particular software domains. Rosson and Alpert (1990) suggest that OO may be especially valuable in new domains or when practiced by relatively experienced designers. In contrast, in the procedural paradigm problem decomposition is driven by generic programming constructs and specialised design knowledge. In terms of the concepts of problem and solution spaces introduced by Kant and Newell (1984), this implies that reasoning in the problem space is separate from reasoning in the software solution space; the objects are considered, but remain implicit.

**3.2    Claims about reusability**

It has been asserted that the OO paradigm promotes reuse of software because the code is encapsulated into objects and the internal details of each object are hidden. The claim about reuse rests on an argument that the hierarchies which form the model of classes are well-suited for reuse (Johnson and Foote, 1988). The programmer needs only to adopt a hierarchy appropriate for the domain of the problem, and then provide the specialisation needed for a particular problem by adding new low-level classes. Thus, much of the needed structure and functionality already present in the higher levels of the class hierarchy is automatically reused by inheritance.

**3.3    Claims about OOD at the software design team level**

Advocates of OOD assert that the OO paradigm helps to overcome problems encountered at the software design team level, compared to traditional paradigms. These problems concern communication, coordination, and the capture and use of application (or problem) domain knowledge (Krasner, Curtis and Iscoe, 1987; Curtis and Walz, 1990; Herbsleb & Kuwana, 1993).

Communication and coordination is enhanced because a single representation, based on the problem domain, underlies all stages of development. Also, communication between designers and clients should be facilitated because the solution is structured around the application domain objects, and is thus easily understood by specialists in the domain (Herbsleb, Klein, Olson, Brunner, Olson & Harding, 1995). OO design should also help to propagate application domain knowledge through the design team, because the shared representation has its basis in the concepts and vocabulary of the application domain.

## 4. OO design by individuals

In this section, we present results of empirical studies of the design activity of individuals. First we present results concerning OO novices. Then we analyse the design activity of OO experts and make a comparison between paradigms. This is followed by a discussion of a typology of problems for OO design. In the final section, we assess empirically the claims about naturalness and ease of OO design.

### 4.1 Design activity of novice OO designers

First, it should be remarked that, in all the empirical studies of novice OO designers, as far as we know, participants had previous experience with the procedural paradigm[1]. Therefore the claims about naturalness of OO design are still to be assessed for real novices. We will see in this section that the previous experience in procedural paradigm causes interference when learning an OOP language.

#### 4.1.1 Difficulties in the process of classes creation

Empirical studies show that novice OO designers tend to identify first domain entities, then classes, and then methods. They follow a method learned in their programming courses. They seem

---

[1] In Rist's study (1996), two subjects started programming by learning an OOP language. However, the analysis was not focused on the specificities of these subjects' activity. In one of the experiments conducted by Chatel (1997), several subjects were real novices. Some preliminary results will be reported in 4.1.1.

to hope that the classes created in this way will be useful later on (Lee & Pennington, 1994; Pennington, Lee & Rehder, 1995)

Novices behave this way not just because they follow a method learned in class but also because of environment constraints, i.e., order constraints which entail premature commitment (Green, 1989). Although this last aspect could not be illustrated in Pennington study, as the novices used only pen and paper, it was illustrated in Détienne's study (1995) in which programmers used an OO environment, the $CO_2$ system. It was observed that the novices, like the experts, tried to define the class model before they implemented the methods in detail. This behaviour was driven by specific characteristics of the version of the $CO_2$ system which was assessed. It was not possible to use an object in a method body if this object had not been completely specified beforehand. The results showed that all classes created in the earliest stage of design were not useful later on. Revisions such as deletion and addition of classes were made later on.

Several studies (Détienne, 1990a; 1990b; 1995; Lee & Pennington, 1994; Pennington, Lee & Rehder, 1995) attest of the difficulties encountered by novices in the process of class creation. In Pennington et al.'s study, it was found that novices spent a considerable amount of time creating and abandoning domain entities before they actually started defining the design objects. The activity of finding classes of objects consumed novices' attention, and they gave no attention to functionality until late in the design activity. In fact, novices delayed consideration of goals and the definition of methods until late in the design. The same process of creating and abandoning classes was observed in Détienne's studies.

Chatel (1997) observed that real novices, without previous experience in another programming paradigm, delayed consideration of goals. They start by creating classes with some typical functions associated to each class. This suggests that structural schemas representing typical functions or roles associated to classes are acquired relatively early in the acquisition process. This results is similar with what was found in procedural design (Rist, 1989). However, Chatel also shows that novices spend much time defining and revising classes.

#### 4.1.2 Difficulties in articulating declarative and procedural aspects of the solution

One of the main difficulties experienced by novices is the articulation between declarative and procedural aspects of the solution, e.g., the hierarchy of classes and the main procedure (Détienne, 1990a; 1990b; 1995). Based on knowledge in the problem domain, novices identify the structural properties and organisation of classes. Based on knowledge in the programming domain, they identify procedural aspects of the solution. As a first step, they often construct a procedure to deal with the main goal (a complex plan) as well as procedures for performing typical roles (simple plans), e.g., initialisation. The complex plan represents a large procedure, not necessarily expressed at the code level yet, and not yet decomposed into individual functions which could be associated to classes.

At this phase in the design process, there is only slight integration between the declarative and the procedural aspects of the solution. Our results (Détienne, 1995) tend to show that novice OO designers describe objects and actions separately in their first draft of the solution more often than do experienced OO designers, especially for a declarative problem. Associating declarative and procedural aspects implies decomposing the large procedure into smaller functional units. However, some novices do not do so. Instead, they associate the procedure as a whole to a single class (Détienne, 1990b).

When decomposing the procedure into functional units and associating them to classes, the decomposition process follows either a procedure-centred strategy or a function-centred strategy. This process was analysed, in our study (Détienne, 1995), by identifying the relationship between methods defined sequentially. It was found that for novice OO designers methods are defined both following their execution order (i.e. a procedure-centred strategy) and according to their functional similarity (i.e. a function-centred strategy) regardless of the objects to which they are associated. It is striking that in both cases objects are secondary entities to be associated with procedure units or functions.

When they refine the functions by writing the code of methods, novice OO designers are able to refine some characteristics of the structure of objects already defined at higher levels of abstraction, e.g., they add attributes in a class or add parameters in a method signature. However, they also make more drastic changes to the declarative model. They add new classes and new methods and move methods from one class to another. Not surprisingly, more modifications are made by novice OO designers than by experienced OO designers. Whereas this difference is not significant when we consider the total number of modification, it is significant when we consider only the methods modifications, e.g., adding/modifying/removing a method and moving a method from one class to another (Détienne, 1995). These modifications concern the definition of a complex plan and the association of parts of this plan with objects.

This result exhibits the difficulties novices OO designers encounter in associating objects with procedures. The lack of appropriate knowledge in OOP and the transfer of inappropriate plans from procedural languages does not allow the construction of a correct complex plan and the allocation of actions of this plan among classes. When they construct a more detailed representation of the procedure (developed in the procedural part), they revise the static aspects of the global solution, i.e., parts composing a complex plan or the association between the actions of this plan and the objects. This suggests that constructing a representation of some procedural aspects of the solution precedes the construction of more declarative aspects of the solution, e.g., the objects. This result provides support to the hypothesis made in cognitive ergonomic, that: (1) knowledge is organised on the basis of goals and procedures and not on the basis of relational properties of the objects and, (2) the procedures are properties of objects and they form the basis for objects categorisation (Richard, 1996)**.**

Novices compile their programs frequently (Détienne, 1993). They tend to write pieces of code corresponding to one or several methods then compile immediatly. Thus, they use a generate/test and debug strategy, the evaluated units being one method or several methods.

4.1.3  Misconceptions

Novices OO designers have misconceptions about some fundamental OO concepts (Détienne, 1990a; 1990b), in particular the class concept and the inheritance property. Novices OO designers tend to conceive a class as a set of objects. Therefore, they attribute set properties to the class concept. This misconception is revealed by errors of novices. For example, when they define a function which processes a set of objects of classX, they place this function in classX instead of creating a classY whose type would be "set of objects of the classX" and placing the function in classY.

Novices tend to think that instances of classes are created by default. This misconception is also revealed by errors. Novices tend to use instances of classes without creating them beforehand. This type of error is similar to forgetting to initialise variables in the procedural paradigm. Initialising a variable or creating an object is a type of prerequisite which is not necessary in the world situation and so is not part of the knowledge transferred from the real world. In general, the novice learns by analogy between the real world source situation and the OO target situation. However, here there are preconditions to be satisfied in the target situation which are automatically[2] satisfied in the source situation (Hoc & Nguyen, 1990).

Novices have also misconceptions about inheritance. They try to use this property as often as they can, and they often do it inappropriately. In particular, they tend to use the abstraction hierarchy to express a composition. There are two aspects of inheritance: the inheritance of static characteristics of a class, i.e., the type of the class, and the inheritance of functional characteristics, i.e., the methods of a class. Our observations tend to show that, while novices have difficulties in using the inheritance property of static characteristics, as shown in the misconception explained before, they have even more difficulties in using the inheritance of functionality.

4.1.4 <u>Transfer when shifting from traditional design to OO design</u>

---

[2] In the real world, objects are usually created beforehand by somebody else. When using an object a person does not have to worry about creating it.

Transfer effects between languages and design approaches have been well documented for procedural programmers learning another procedural/functional language or a declarative language (Siddiqi, Osborne, Roast & Khazaei, 1996; Scholtz & Wiedenbeck, 1990, 1993; Wu & Anderson, 1991). For OOD, it has been claimed that previous knowledge of the problem domain should help in analyzing the problem and decomposing and structuring the solution. However, most beginners in OO design have previous experience with other design approaches such as procedural design, and there is strong evidence of knowledge transfer not only from the problem domain but also from other languages and from other design approaches. Thus, one issue of concern is the transfer of knowledge from traditional design and the effect of this transfer when shifting from traditional design to OO design. Some studies have dealt with this transfer issue (Détienne, 1990a; 1990b; 1995; Pennington, Lee & Rehder, 1995) and also with the assessment of various transfer indicators (Chatel, Détienne & Borne, 1992).

It has been shown that novices OO designers transfer knowledge from non-OO design methodologies. For example, in one of our own studies (Détienne, 1990a; 1990b), we observed that a novice OO designer with previous experience in procedural programming used attributes of type "number" in each class to link together objects, instead of using the is-part-of relationship. From the verbal protocol, it was evident that this subject constructed a solution using elements of a relational approach to data base management. According to this methodology, different objects have a number which is used as a cue to link together objects and to facilitate the search in a data base. Evoking this schematic knowledge, the subject added an attribute of type "number" to each class he had constructed previously. Then he constructed a kind of "flat" structure of classes, without using the "is-part-of" relationship to link together classes. Obviously the solution produced by this subject did not conform to principles of OOP.

Other evidence of transfer from procedural languages and design is provided by Pennington et al. (1995). They observed that OO novices retain a few procedural features in their design, such as the use of input/output objects that were extensively interconnected to every other object in the design. In one of our own studies (Détienne, 1990a; 1990b), we also observed such transfer effects. For example, a novice OO designer added a parameter "type of object" in a class which allowed him to do different processing (method calls) according to the value taken by this parameter in a "case of" or "ifs" structure. This is a typical solution used in classical procedural languages. In doing so, the novice failed to take into account the functionalities of OOP. By using late-binding and inheritance, he could have allowed the system to decide during execution which kind of object was being processed. In this way the appropriate method could have been called without using a "type" parameter and a selection structure.

We have been particularly interested in comparing the errors made in declarative problems and procedural problems (Détienne, 1995). In declarative problems, the program structure is strongly constrained by the data structure: the representation of this structure guides problem development. In procedural problems, the program structure is strongly constrained by the procedure structure: the representation of this structure guides program development (Hoc, 1983). Thus, the nature of what is transferred should be different depending on the problem type.

In our experiment, we used two management problems: (1) a library management problem that was a slightly modified version of a problem classified by designers in a previous experiment (Hoc, 1983) as a procedural problem and (2) a financial management problem classified as a declarative problem. Our results (Détienne, 1995) tend to show that the definition of the structure of classes is more error prone for a declarative problem. This suggests that when a strong data structure known from procedural languages is transferred, the resulting representation includes unnecessary classes and/or objects structured in a way more appropriate to procedural programming. For the procedural problem, we more often found errors of misplaced methods. This suggests that when a strong procedure structure known from procedural languages is transferred, the structure of this procedure is developed as a complex plan. Linking parts of this plan with classes is error prone.

### 4.2 Comparing OO design with procedural design

In this section, we examine the processes involved in designing in the OO paradigm and the procedural paradigm. We will refer to results of comparative studies and also to studies that analyse the cognitive processes of procedural designers, on the one hand, and OO designers, on the other hand. It should be noted that the latter comparison is more risky as it is more difficult to control the effect of other factors besides the paradigm itself.

4.2.1   Methods for comparing design across paradigms

Comparing design in different paradigms raises specific methodological problems. Some authors (Lee & Pennington, 1994; Pennington, Lee & Rehder, 1995; Rist, 1996) have made an important contribution by developing methodologies, independent of the design paradigm, for the description and evaluation of final designs and for analysing design processes.

The methodology developed in the Pennington study is based on the design methodology of Rumbaugh, Blaha, Premerlani, Eddy and Lorensen (1991). It allows the comparison of the final design across different design paradigms. It examines how design proceeds through three types of structure: functional, procedural and object structure. It also allows measurement of the completeness of the final design from each of these three different views.

Rist has developed a powerful tool, called the system structure, to formalise design solutions. Three types of links are represented across goals and objects: control flow, data flow, and sequence. This tool can be used to analyse design activity by showing the order in which nodes are added to the system. Three dimensions of design capture the order in which nodes are added to the system structure: link type (goal versus class), level (top-down versus bottom-up) and direction (forward versus backward).

Whereas several authors used a paradigm for comparing OO design with procedural design, we can regret that, to date, the methodologies developed by Pennington and by Rist have been used only in these authors' own studies.

4.2.2  Ease of mapping between the problem space and the design solution space

Several comparative studies (Kim & Lerch, 1992; Boehm-Davis & Ross, 1992; Lee & Pennington, 1994) show that OO design tends to be easier and faster than procedural design. With regard to the claims made about naturalness and ease of design in the OO paradigm, an important issue is whether the problem decomposition is driven by problem domain entities. In several comparative studies ( Lee & Pennington, 1994; Pennington, Lee & Rehder, 1995; Rosson & Gold, 1989) it has been shown that problem decomposition is driven by problem domain entities for OO expert designers whereas for procedural designers the decomposition is driven by generic programming constructs and specialised design knowledge.

In Pennington's studies ( Lee & Pennington, 1994; Pennington, Lee & Rehder, 1995) it was found that expert OO designers analyse a situation through its objects. Compared to procedural design, less time is spent analysing the problem situation and more time is spent describing objects and their relationships. Differences between procedural and OO experts are noted mainly in the way in which they analyse the task domain. Expert OO designers spend as much time analysing the domain as expert procedural designers but they accomplish this through the creation of classes of objects rather than in general terms. Pennington et al. interpret this result as supporting the claim that OO designers analyze situations through objects and their interrelations.

The same result was found in Rosson and Gold (1989). OO designers use knowledge of how problems entities operate and interact to reason about design objects. On the basis of their observations the authors suggest that much of OOD in its earlier phases is driven by an understanding of the problem itself rather than by specialised design knowledge**.** On the contrary, the solutions of procedural designers are structured by generic programming constructs rather than by entities of the problem domain.

Furthermore, it has been found that compared to procedural designers, OO designers produce more similar final solutions with similar objects and methods (Boehm-Davis & Ross, 1992; Lee & Pennington, 1994). This similarity could be attributed to a close mapping of the problem domain.

This tends to support the claim that OO designers are able to create designs that map more closely to the problem domain (Rosson & Gold, 1989).

4.2.3 Design strategies

In OO design, two classes of strategies may guide the design activity: strategies based on representations exhibiting static characteristics and strategies based on representations exhibiting dynamic characteristics. We will refer to these two classes of strategies by the terms "declarative plan" and "procedural plan".

When the plan[3] which guides the design activity is declarative, static characteristics, such as objects and typical functions (roles), guide solution development. The use of two strategies suggest the involvement of such a plan: strategies based on typical functions or strategies based on objects. These strategies (Chatel & Détienne, 1994; 1996) are:

-function-centred strategy: the functions are prominent in the representation guiding the design activity, and objects are subordinate to functions. The subjects follow plans in which typical functions (or roles) are central. They develop one function for several objects, then another function for several objects, and so on. Examples of functions are reading access, writing access, initialisation, printing, etc.

-object-centred strategy: the objects are prominent in the representation guiding the design activity, and functions as well as procedures are subordinate to objects. This means that objects are central. The subjects tend to follow a plan in which several typical functions as well as typical calling structures are associated to a generic object. The subjects develop several typical functions for one objects then the same functions for another object and so on. It may also happen that the subjects develop several typical functions by following their calling links in objectA then develop several similar functions by following their calling structure again in objectB, and so on.

---

[3] Here the term "plan" refers to the mental representation which guides the design and does not mean a solution plan.

When the plan which guides the design activity is <u>procedural</u>, dynamic characteristics, such as actions organised in the temporal order of their execution, guide solution development. The use of a strategy based on procedures (Chatel & Détienne, 1994; 1996) suggests the involvement of such a plan:

-<u>procedure-centred strategy</u>: the representation of the procedure guides the design activity. The subjects tend to follow plans in which methods implied in a procedure are organised according to their calling structure. In OO programs such a procedure is a complex plan which is delocalised. Actions of a complex plan correspond to single methods which are associated to different objects. The client-server relationships, e.g. the calling structure, provide the links between those actions. According to the procedure-centred strategy, the subjects develop in a row several methods implied in a procedure for performing one main goal of the problem, even if these methods are associated to different objects. They follow the message passing (or calling) structure.

These strategies have been observed with the CO2 language in Détienne's study (1995) and with Smalltalk in Chatel & Détienne's study (1994). Similar search strategies have been described in a study conducted by Rist (1996): a design strategy based on roles, a design strategy based on goals, and a design strategy based on objects.

- a <u>design strategy based on roles</u> expands one role at a time. Similar roles exist across goals and/or objects. In Chatel and Détienne terminology, this is a function-centred strategy.

- a <u>design strategy based on goals</u> expands one goal at a time, varying the role and the object. In Chatel and Détienne terminology, this is a procedure-centred strategy.

- a <u>design strategy based on objects</u> expands one object at a time, varying the roles and goals relevant to those objects. In Chatel and Détienne terminology, this is a object-centred strategy.

These two kinds of plans, declarative and procedural, are also used in procedural design. The concept of declarative versus procedural plans is related to Hoc's (1983) declarative/procedural problem dimension. A problem is procedural if the representation of the procedure structure guides solution development. A problem is declarative if the representation of the data structure guides solution development.

An important issue is to determine what conditions trigger the use of one kind of plan rather than another in OO design. It seems that the use of these strategies may be a function of the expertise and of the type of problems. Concerning the effect of expertise, Détienne (1995) observed that novice OO designers more often used a procedural plan, i.e., a procedure-centred strategy, than experienced OO designers. Experienced OO designers tended to use mostly a declarative plan based on the objects, i.e., an object-centred strategy.

In Chatel and Détienne (1996) expert OO designers tended to use one strategy rather than another depending on the type of problem. It was found that, for problems which had some procedural characteristics, in particular a flat structure of objects with horizontal communication between objects, even experts OO designers tended to use a procedure-centred strategy. For problems which had a hierachical structure, experts tended to use either an object-centred strategy or a function-centred strategy. This issue of typology of problems for OOD is discussed in Section 4.3.

By contrast, Rist (1996) observed that the strategies of experienced OO designers were based most commonly on goals, i.e., a procedure-centred strategy, and less commonly on objects, i.e., an object-centred strategy, whatever the type of problems (data flow versus control flow problems). Increased difficulty led to an increase in goal-based design (procedure-centred strategy). Here difficulty[4] was evaluated on the basis of the length of the program and the time spent to develop it.

The procedure-centred strategy is the most common strategy used by novices in the Détienne study, the most common strategy used by experienced programmers in the Rist study and it is used by experts only for procedural-like problems in the Chatel and Détienne study. An explanation of these contradictory results could lie in the way in which subjects are categorised as experienced designers or novices in these studies. The so called "experienced designers" were students in the Rist study,

---

[4] This way to assess problem difficulty reflects the quantity of information to manage when solving the problem rather than the complexity or familiarity of the problem.

whereas they were professional in the Détienne study and in the Chatel and Détienne study. This may explain why experienced designers of the Rist study are closer to novices designers than to experienced designers in the Détienne study.

More generally, comparing results of different studies by subject expertise is often difficult because different authors categorise subjects differently. Even the number of years or experience or the number of known languages may not be a good way to distinguish between different levels of expertise. A good dicussion of this issue is presented in Sonnentag (1996) who defends the idea of using the judgement of subject's peers as an additional way to distinguish between levels of expertise.

Each of these studies shows dominant strategies for groups of designers but also shows that individual designers shift between strategies when solving a problem. Rist (1996) discusses these results as showing that design is an active, local and opportunistic activity.

Another strategy, more related to the design evaluation, is the use of <u>mental simulation</u>. The use of this strategy has been widely documented in studies on procedural design (see, for example, Adelson and Soloway, 1988; Guindon, 1990a). On this point, there are contradictory results in the studies comparing OO and procedural design. In Lee and Pennington (1994), it is found that OO designers spend a greater proportion of time evaluating their designs through mental simulation than do procedural experts. On the contrary, Kim and Lerch (1992) observed less mental simulation in OO design compared to procedural design. It is worth noting that in this latter study there was only one OO designer while in the former study there were four OO designers. This should warn us to be cautious with hasty generalisations.

4.2.4   <u>Organisation of the design activity</u>

A higher level of analysis is the level of the organisation of the design activity which is a meta-level to the strategy level. At this meta-level, empirical studies of software design have highlighted the opportunistic nature of the organisation of the design activity (Guindon, Krasner &

Curtis, 1987; Guindon, 1990a; Visser, 1987; 1994). Even though OO designers may organise their activity opportunistically, as observed in Rist (1996), it is still necessary to assess whether OO design is relatively more or less opportunistic than procedural design. On this point, two comparative studies provide us with contradictory results.

Pennington et al. (Lee & Pennington, 1994; Pennington, Lee & Rehder, 1995) found more opportunistic design among procedural designers and more top-down, breadth-first design with less opportunistic behaviour among OO designers. On the contrary, Brangier and Bobiller-Chaumon (1995) found more opportunistically organised design among OO designers and more hierarchically organised design among procedural designers.

An explanation for these contradictory results may be found in the different experimental conditions of these two studies: use of pen and pencil in Pennington et al.'s study versus use of programming environments in Brangier and Bobiller-Chaumon's study (C++ environment versus Pacbase for Cobol). Thus, it is not clear in fact what is measured: the effect of the environment or the effect of the paradigm.

Some OO environments allow opportunistic design, for example, the Smalltalk environment as shown in an empirical study on reuse (Rosson & Carroll, 1993) and in the claims analysis of the Smalltalk browser (Bellamy, 1994). However, we argue that OO environments do not necessarily allow or support opportunistic design. For example, the $CO_2$ environment, at least in an early version that we assessed (Détienne, 1990a; 1990b), constrained designers mostly to top-down design: creation of classes in a top-down manner and creation of a method's signature in the class description before defining the method's code.

Empirical studies of software design have also highlighted the use of kernel concepts by expert procedural designers (Guindon, 1990b). Such kernels seem to be used by OO designers as well: they use plan schema or abstract design schema as an abstract description of the design from the beginning of the design sessions (Lee & Pennington, 1994; Pennington, Lee & Rehder, 1995; Rist, 1996). A question is the structure of kernel concepts in these paradigms.

### 4.3 Toward a typology of problems for OOD

Some problems may be better suited to a non-OO solution because the program structure created by a language may facilitate or obscure the clear expression of a solution (Petre, 1990). Thus, it is important to look for a typology of design problems. The critical question is which dimensions of problems are relevant for determining the correspondence between the problem and the design approach.

Hoc (1981; 1983) proposed a framework for classifying problems and strategies. Two dimensions are distinguished: procedural versus declarative, and prospective versus retrospective. In procedural problems, the program structure is strongly constrained by the procedure structure. The representation of this structure guides solution development. In declarative problems, the program structure is strongly constrained by the data structure. The representation of this structure guides solution development. Furthermore, the solution may be developed in a prospective (forward) way, e.g., when the data structure of the input is strong, or in a retrospective (backward) way, e.g., when the data structure of the output is strong. These distinctions have been made in the context of procedural design. It is important to analyse whether the same dimensions are relevant for OOD and the conditions which influence the use of one strategy rather than another.

A dimension which seemed, a priori, more relevant for the OO approach, is the distinction between declarative versus procedural problems. In previous work (Détienne, 1990), we hypothesized that, in the OO paradigm, declarative problems would be easier to solve than procedural problems. This hypothesis was not confirmed by our results. For example, we observed an equivalent proportion of revisions whatever the problem type. Furthermore, our study showed that experienced OO designers tended to use a declarative plan whatever the problem type. Rist (1996) found an effect of the problem type only in early planning. Whereas early planning was done following the control flow for most experienced designers, data flow was followed only for the data flow (or declarative) problems. Thus, this problem dimension does not seem to influence greatly the design activity. A question is whether or not another dimension of the problem may influence the choice of design strategies in OOD.

Another dimension was proposed by Chatel and Détienne (1996). This new dimension characterises not only the data structure (or objects) and the procedures, but also the way they are associated. The structure of the solution and the way objects communicate within this structure is an important feature of solutions in OOD.

For problems with a hierarchical structure of classes with vertical communications between objects, we observed (Chatel & Détienne, 1996) that experts OO designers used a declarative plan. Static characteristics such as objects and typical functions guided solution development. For problems with a flat structure of classes with horizontal communications between objects, i.e., more procedural-like solutions, we observed that experts OO designers used a procedural plan. Dynamic characteristics of the procedure guided solution development.

Of course, it could be argued that the OO paradigm encourages the development of hierarchical solution structures rather than flat solution structures. However, the point is that, for any large scale software development, we could distinguish parts of the solution with a hierarchical structure from parts with a relatively flat structure. This being the case, it is likely that various design strategies, and more generally various kinds of plan (declarative versus procedural), may be used for developing parts which have different structures.

Another problem dimension relevant to OO design is discussed by Herbsleb et al. (Herbsleb, Klein, Olson, Brunner, Olson & Harding; 1995). It characterises whether or not well understood abstractions have already been constructed in the task domain. These authors distinguish problems for which domain knowledge is known and has just to be captured from problems for which domain knowledge has to be invented. Whereas OO design may be relevant and easy for the former type of problem, what happens with the latter type of problem is still not clear.

**4.4     Assessing the claims about naturalness and ease of OO design**

We will assess the claims on naturalness and ease of OO design first for novices and then for experts. The literature on novice OO designers shows that they have difficulties in the process of class creation and in articulating the declarative and procedural aspects of their solutions. These results do not support the claims about naturalness and ease of OO design. First they show that the identification of objects is not an easy process. On the basis of knowledge in the problem domain, entities are identified as objects but these entities are not necessarily useful in the design solution. The mapping between the problem domain and the programming domain is not straightforward. The analysis of the problem domain is not sufficient to structure the solution in terms of objects. The results show that the novices need to construct a representation of the procedural aspects of the solution in order to refine, evaluate and revise this decomposition into classes with actions associated. Our interpretation of these results is that (1) knowledge is organised in terms of goals and procedures and not in terms of relational properties of the objects and (2) procedures are properties of objects and they form the basis for objects categorisation.

The literature on novice OO designers also shows some negative effects of knowledge transfer from other paradigms. This kind of result was anticipated by some advocates of OO paradigm. For example, Rosson and Alpert (1990) noted that "However, there exist many designers already trained in traditional design methodologies, and we know little about how best to ease their transitions to the object-oriented approach."

The literature on expert OO designers apparently supports the claims about naturalness and ease of OO design. It shows that decomposing the problem is driven by problem entities for expert OO designers, whereas for procedural designers the decomposition is driven by generic programming constructs and specialised design knowledge. Expert OO designers spend as much time analysing the domain as expert procedural designers but they accomplish this through the creation of classes of objects rather than in general terms. This supports the claim that OO designers analyze the situation through the objects and their relationships. Furthermore, results support the claim that OO designers are able to create designs that map more closely to the problem domain.

Putting all these results together makes an odd picture. If OO design decomposition is mostly driven by domain knowledge then the biggest benefit of this approach should be observed in novices. However, this result is not found. Two explanations cand be proposed. First, it is possible that the results found for novices reflect mostly the negative effect of knowledge transfer from other paradigms. In this regard, studies with real novices would be extremely useful. As noted by Rosson and Alpert (1990), an interesting test would compare novice programmers learning to design in the procedural vs object-oriented paradigms.

A second explanation is based on the nature of knowledge constructed through experience with the OO paradigm. Détienne (1995) suggests that different schemas, depending on subjects' language experience, are used for planning activities. Schemas related to procedural languages group actions in execution order. Schemas related to OO languages integrate actions and objects, with actions organized around objects. The latter type of schemas may be developed through practice with OOP languages and is more adapted to the constraints that must be taken into account when designing with this kind of language (e.g., making explicit the links between objects and actions).

Furthermore, the results show that expert OO designers may shift from an object-centred to a procedure-centred strategy when developing their solution. This result shows that the designers shift between the object view and the procedure view and that both views can be considered as important entities in the design process. This shift may also be systematic for certain types of problems as shown previously in the discussion of the typology of problems. These last remarks are consistent with Rist's theoretical approach to OO design (Rist, 1996). He defends the idea that plans and objects are orthogonal which means that they are both first-class entities conceptually: they present different but valid views of a system, are both important, and can both be used to drive the design activity.

## 5. **OO software reuse**

The OO design approach supports software reuse through the mechanisms of abstraction, encapsulation and inheritance. We will first compare reuse in the OO and the procedural paradigm.

Then we will discuss results indirectly assessing the potential for reuse in OOD. The next topic concerns the strategies and processes involved in OO software reuse. Finally, we will assess empirically the claims on OO reusability.

## 5.1 Reuse in the OO versus procedural paradigms

As far as we know, there is only one study comparing reuse in the OO and the procedural paradigm. An exploratory study by Lewis, Henry, Kafura, and Schulman (1991) presented evidence of increased productivity in OO programming, much of which they attributed to reuse. The main conclusions of the authors (p 195) are:

"-the OO paradigm substantially improves productivity, although a significant part of this improvement is due to the effect of reuse;

-Software reuse improves productivity no matter which language paradigm is used;

-language differences are far more important when programmers reuse than when they do not, and

-the OO paradigm has a particular affinity to the reuse process."

Some studies have assessed the modifiability of OO programs. Considering that modifying a program is part of the reuse activity[5], the results of these studies may shed light on the reusability issue. In a study on OO system maintenance (van Hillegersberg, Kumar and Welke, 1994), it was found that programmers with experience in structured development but with low experience in OO development had trouble understanding and maintaining an OO system. Henry and Humphrey (1993) found OO programs to be more easily modifiable than procedural programs for undergraduate students. Boehm-Davis, Holt, and Schultz (1992) found the reverse result for student programmers when they made complex modifications. They found no significant effect of the program structure, in particular functional versus OO, for professional programmers. However, in this latter study, the OO programs were constructed in Pascal and did not have all the OO properties, in particular, inheritance was missing.

---

[5] Three processes, traditionnally studied in analogical reasoning studies, are involved in the reuse activity: retrieval, mapping and adaptation/modification.

**5.2    Potential of OO software reuse**

A way to assess the potential for reuse in OO design is to analyse the consistency of final designs and representations constructed by designers. According to Lee and Pennington (1994), more similar decompositions are produced by OO designers than by procedural designers. In their study, OO designers produced more closely matched designs with similar objects and methods, whereas procedural designers produced less closely matched designs with different types of data abstractions and procedural decompositions. Similarly, Boehm-Davis & Ross (1992) found more consistency in solutions produced following an OO approach than a functional decomposition approach.

This consistency may be tied to the shared problem domain knowledge of the designers (Dvorak & Moher, 1991). In a study on class hierarchy construction, these authors found that differences in domain experience resulted in qualitative differences in the approach to the problem and impacted inter-subject agreement on the structure of the resultant hierarchies.

This consistency decreases as the hierarchy becomes deeper. In an experiment in which subjects were to construct a class hierarchy using class specifications, Dvorak (1994) found that, the deeper the hierarchy is, the less agreement there is between subjects about the class's placement in the hierarchy. This author explains this by using the concept of <u>conceptual entropy</u>: "Conceptual entropy is manifested by increasing conceptual inconsistencies as we travel down the hierarchy. That is, the deeper the level of hierarchy, the greater the probability that a subclass will not consistently extend and/or specialise the concept of its superclass." This concept may explain why, as the hierarchy becomes deeper, the performance of subjects in a maintenance task deteriorates, as found in a recent study comparing 3 levels of inheritance versus 5 levels of inheritance (Daly, Brooks, Miller, Roper & Wood, 1996).

Reuse by inheritance is dependent on the organisation of objects. Reuse in OOP is based on the assumption that the representation of objects is most important and corresponds to the deep structure of programs. However, empirical studies have found that the object view is not the most dominant view of the code (Davies, Gilmore & Green, 1995; Chatel & Détienne, 1994). These authors found the following results:

-experts may organise their thinking about programs according to three views: functional, procedural and object

-the functional properties and message passing relationships are more important for experts than for novices. Experts organise their thinking about programs around the algorithmic structure

-the relationships of objects are more important for novices than for experts. Novices organise their thinking about programs according to the problem domain

These results challenge the idea that the objects form the deep structure of OO programs. In a general way it is likely that different tasks require different perspectives on the code, not all of which are object based.

### 5.3 Reuse activity

Supporting reuse activity has become a big challenge in software engineering. At the same time, empirical work on design with reuse activity has been conducted in recent years, particularly on reuse in the OO context. A first general remark is that reusing through built-in classes is not spontaneous. Expert OO designers utilise reuse through built-in-classes whereas the novices do not spontaneously take advantage of built-in classes (Lee & Pennington, 1994). Second, contrary to the claims about how OO languages might facilitate reuse, Lange and Moher (1989) and Détienne (1991) found that reuse efforts did not centre on reuse by specialising the inheritance hierarchy but rather on literal copying and modification of code. Third, reuse involves some particular expertise different from expertise in the programming domain and expertise in the problem (task) domain. Programmers who are experienced in programming and familiar with the application domain, but novices in the reuse task, have difficulties in selecting, and modifying a reusable component. Their criteria for choosing between several reusable components are inadequate (Rouet, Deleuze-Dordron & Bisseret, 1995; Woodfield, Embley, & Scott, 1987). Although a greater potential for reuse may exist in OO programming, it may not be enough to promote reuse without the development of support tools and reuse methods/training.

5.3.1  Cognitive typology of reuse processes

From the software engineer's point of view, a classical typology of the reuse situation is made in terms of the type of reusable components (e. g. code, specialisable components, class reuse by inheritance). From a cognitive point of view, we have proposed a typology of reuse processes[6]. The typology is based on the cognitive status of the reused component, which depends on the goal of the designer. This typology is orthogonal to the previous typology of components.

Three situations (Burkhardt & Détienne, 1994; 1995a; Détienne, 1996) can be distinguished: (1) the designer analyses the software problem and then the retrieval of a reusable component allows the addition of specifications to the problem at hand, the addition of constraints or the abstraction of constraints; (2) the designer is looking for a solution and then the retrieval of a reusable component allows the evocation of alternative solutions, the evocation of criteria to evaluate a tentative solution, the construction of a plan which guides the problem solving process, or the revision of the goal structure; (3) the designer has already chosen and begun implementing a solution and is looking for a reusable component to avoid the coding of the chosen solution[7]. Thus, at one extreme of this continuum, designers may look for a model which guides their designs and, at the other extreme, they may look for a pluggable component which allows the implementation of a chosen solution without having to code it.

---

[6] We think that this cognitive typology of reuse situations can be related to a typology of documenting processes. A theoretical framework for documenting reusable components should be based both on our typology of reuse processes and on our typology of documenting processes (Détienne, Rouet, Burkhardt & Deleuze-Dordron, 1996) as well as on the idea of using free annotations (Green, Gilmore, Blumenthal, Davies & Winder, 1992).

[7] Considering separately the phases of analysis, problem solving and implementation helps us to emphasize the different nature of the reuse processes which are involved in these phases. However, these phases are not striclty separate and there are interactions between them as shown in empirical studies on opportunistic design (Guindon, 1990; Visser, 1994).

### 5.3.2 Enrichment of the representation versus lowering of the level of control of the activity

Our typology of reuse processes allows us to explain apparently contradictory results of empirical studies on reuse, results which seem not to be specific to reuse in the OO context, by the way. Some results are that the effect of the reuse processes is an <u>enrichment of the representations</u> constructed during design (Burkhardt & Détienne, 1995b; De Vries, 1993; Rosson & Carrol, 1993), e.g. by the inference of new constraints, new goals. Other results are that the effect of the reuse processes is the <u>lowering of the level of control of the activity</u>, in particular by the use of test/debug strategies and comprehension avoidance strategies (Lange & Moher, 1989; Détienne, 1991). We defend the idea that the former type of results concern reuse processes involved during the analysis and problem solving phases whereas the latter type of results concern reuse processes involved during the implementation phase.

The effect of the reuse processes may be an enrichment of the representations constructed during the problem analysis and problem solving phases. Burkhardt and Détienne (1995) show that evoking a reusable component may allow the addition of constraints, evaluation criteria, new goals. Rosson and Carrol (1993) note that sometimes the borrowed code is not directly reusable itself but rather is used more as a functional specification.

Reuse results in the lowering of the level of control of the activity during the implementation phase. We refer to the hierarchy of levels of control developed by Rasmussen and Lind (1982). These authors distinguish between automatic activities, activities based on rules, and activities which involve high-level knowledge. The lowering of the level of control of the activity consists in switching from activities which involve high-level knowledge, e.g. problem solving activities, to activities based on rules and automatic activities, e.g. execution of procedures.

The use of the copy/edit style attests this effect. The copy/edit style does not conform to the style of reuse encouraged for OOP, i.e., reuse by inheritance. It reflects comprehension avoidance of the copied code and use of surface-level features to construct a representation of it (Davies, Gilmore & Green, 1995; Lange & Moher, 1989; Rosson & Carroll, 1993). The designers make "probable" modifications and rely heavily on the debugging tools to evaluate the code. In what we called "new code reuse" (Détienne, 1991), the designer anticipates the reuse of a component in the same program while developing it. In new code resuse designers construct an operative representation of the source as well as a procedure for modifying the code of the source into targets. Designers then execute this procedure to develop targets. In this case also, there is a lowering of the level of control of the activity, which causes errors by propagation of source errors or by omission of changes.

5.3.3   <u>Importance of example and context</u>

Studies show the importance of examples in reuse (Burkhardt & Détienne, 1995; De Vries, 1993) and, in the OO paradigm, the importance of knowledge about the context from which a reusable class comes. In a field study, Rouet, Deleuze-Dordon and Bisseret (1995) found that, for selecting a reusable component in a library, designers were looking for information on the application from which the component was extracted. This contextual information which seems to be highly important is rarely present in the documentation of components because software engineers generally believe that reusable components must be generic and application-independent.

Rosson and Carroll (1993) also note the importance of knowledge about "example application of a target class". They observed that Smalltalk designers reused components indirectly through the reuse of uses. For example, a class might be reused indirectly through the reuse of blocks of code or methods embedded in an example application. The designers relied on code in example applications that provided them with an implicit specification, called the "usage context", for the reuse of the class. The authors observed that, quite often, the usage context was assimilated into the current design project by copying and editing borrowed code. It also sometimes served as a model for the analysis or solution design of the current problem. It is worth noting that the Smalltalk environment supports the identification and reuse of "example usage context" through its sender query which returns a list of methods in which a target message is used.

These authors also observed that Smalltalk programmers sometimes decided to inherit rather than borrow from the example usage context. In this case, they reused more than just the pieces of code involving the target class. They reused the entire application context of the example. This idea has been developed in computer science under the notion of a framework (Fischer, Redmiles, Williams, Puhr, Aoki, and Nakakoji, 1995).

### 5.4 Assessing the claims about OO software reusability

Some results of empirical studies support the claim that the OO paradigm promotes reuse of software. It has been shown that the OO paradigm substantially improves productivity and that a significant part of this improvement is due to the effect of reuse. The potential of OO software reuse is attested by the greater consistency of final designs with this approach compared with a procedural approach. This consistency is tied to the shared problem domain knowledge of the designers.

The style of reuse which is encouraged for OOP is reuse by inheritance. Two kinds of results cast some doubts on the idea that this kind of reuse is easily performed. First, whereas reuse by inheritance is based on the organisation of objects, some results challenge the idea that the objects form the deep structure of OO programs. Three views are important: functional, procedural and object. In a general way, it is likely that different tasks require different perspective on the code, not all of which are object based.

Second, contrary to the claims about how OO languages might facilitate reuse by inheritance, some studies have found that reuse efforts do not centre on reuse by specialising the inheritance hierarchy but on literal copying and modification of code. Furthermore reuse through built-in classes is not used spontaneously by novices. These results suggest that reuse may require a particular kind of expertise and that training and properly documenting components are central issues.

### 6. **OO design by teams**

As far as we know, there are only two studies (Bürkle, Gryczan & Züllighoven, 1995; Herbsleb, Klein, Olson, Brunner, Olson & Harding; 1995) on OOD at the software design team level. These studies have evaluated whether or not problems which have been well documented in procedural design studies also occur in OO development projects. These problems concern communication, coordination, knowledge dissemination and interaction with clients.

### 6.1    Communication and coordination

Herbsleb et al. examined patterns of communication and coordination in design meetings. They compared these patterns between teams using a traditional approach and a team using an OO approach. They found that the major differences associated with OOD were (p273):

"-fewer episodes of clarification in design discussions.

-more episodes of summary and walk-through.

-earlier mentions of criteria in issue discussions.

-more integral role of summary and walk-through in design discussions."

The study found that communication between members of the team was more effective with an OO paradigm than a procedural paradigm: there were fewer episodes of clarification in design discussions. Issues were sometimes addressed by first figuring out what properties a good answer should have, in terms of criteria, before considering alternatives. Criteria were mentioned earlier in issue discussions. OOD helped to focus communication on decisions affecting object interfaces and methods. This suggests that OOD may ease coordination by helping developers work independently and identify what needs to be communicated.

### 6.2    Knowledge dissemination

In the Herbsleb study, the kinds of questions that designers asked in development meetings were analysed and compared between the traditional approach projects and the OO project. The authors distinguish between "what", "why" and "how" questions and between the phase into which the questions are asked, i.e. requirements versus design. They found quite different patterns of questions based on the approach followed by the team. In particular, OO designers asked more questions about design than about requirements and there was an increase in "why" questions. The interpretation given by the authors is that the OO designers were reasoning more deeply about the design. In particular, they sought a more thorough understanding of the underlying issues as revealed by the increase of "why" questions. It is worth noting that this can be a good way to improve the design process.

**6.3 Interaction with clients**

Interaction with clients shows contradictory results. Burkle et al. found that communication between the team and clients was more effective whereas Herbsleb et al. found the contrary result. These latter authors make the distinction between two types of problems (as already discussed in section 4.3): problems for which domain knowledge has to be captured and problems for which domain knowledge has to be invented. They argue that OOD is likely to be particularly useful for the former type of problems. However, it appears that their problem was of the second type, which could explain the difficulties in interaction with clients found in their study.

Herbsleb et al. also discuss the tension between generality and understandability. To use inheritance and reuse effectively, the designers must design classes which are highly abstract. However, the more abstract the classes are, the more difficult they are for users and domain experts to understand.

**6.4    Assessing the claims about OOD at the software design team level**

These results tend to show that the OO paradigm helps to overcome some problems encountered at the software design team level compared with traditional paradigms. Communication between members of the team is more effective. Coordination and Knowledge dissemination are enhanced. Interaction with clients seems to be improved mostly for well-defined problems. However for problems for which domain knowledge has to be invented there is no improvement. This casts doubt on the hypothesis made by Rosson et Alpert (1990) that OO may be especially valuable in new domains.

## 7. Limitations and directions for future work

We would like to highlight some methodological limitations of the studies on OOD. Both controlled studies and field studies are often limited by a low number of subjects. Also, it is often difficult to disentangle the effects of variables such as characteristics of the language, methods, environment, design metaphor, etc. As noted by Curtis (1995, p 341), "we need to distinguish between OO as a computational model and OO as a cluster of mutually supportive development methods, not all of which are strictly tied to the OO computational paradigm".

These studies have shown some benefits of OOD both at the individual level, in particular for expert OO designers, and at the team level. However, we have seen that major problems are encountered when shifting from the procedural to the OO paradigm. With respect to claims on reuse, results of empirical studies show frequent use of the copy/edit style of reuse, which is not the style of reuse encouraged by proponents of OOD.

To go further, it seems important to view the studies presented in this paper and future studies on OO design from the view of methodology: for example, the language and the environment used.

As concern the studies presented in this survey, Pennington's studies and the Rist study were done using paper and pencil. Furthermore, in Pennington's studies, the subjects made high level designs without any particular specified OO language[8] whereas most other studies used a particular OO language (and environment):

- the Smalltalk language and environment in Chatel and Détienne's studies (Chatel and Détienne, 1994; 1996; Chatel, Détienne and Borne, 1992) and in Rosson and Carroll's study (1993),

- the CO2 language and environment in Détienne's studies (1990a; 199Ob; 1991; 1993; 1995),

- the C++ language in Burkhardt and Détienne (1995), in Davies, Gilmore and Green's study (1995) and the C++ language and environment in Brangier and Bobiller's study (1995)

---

[8] Even though the OO languages they were thinking of for implementation was either C++ or Smalltalk.

- the C++ and the Eiffel languages in Rist's study (1996)

One point could be to examine OO languages used according to their similarity with procedural languages. It is possible that the similarity between an OO language and known procedural languages encouraged transfer and intrusions from procedural languages: the C++ and CO2 languages have procedural characteristics (similarity with the C language) that are not present in other OO languages such as Smalltalk. Most studies using either C++ or CO2 provide evidence of negative effects of transfer. However even with Smalltalk, Chatel, Détienne & Borne (1992) found tranfer between paradigms.

Another point could be to analyse languages and environments used in these studies according to various cognitive dimensions (Green,1989) and their influence on the organisation of the activity and on the use of design strategies. For example, it is clear that the CO2 environment and language (Détienne, 1995) have order constraints (an object cannot be used in a method body if it has not been completely specifed before) which force a top down development of the solution. This causes premature commitments and entail many revision afterward. With Smalltalk, authors (Rosson & Carrol, 1993) have illustrated the opportunistic nature of the organisation of the activity which is encouraged by the environment characteristics.

As concern the use of design strategies it is clear that some environment characteristics may encourage the use of one strategy rather than another. For example, if we analyse the strategies used by OO experts, the function-centred strategy was not observed in the Détienne study (1995), at least in the preliminary planning phase, and was observed in the Chatel and Détienne study (1996). A difference between these studies is the OOP language and device used, CO2 in the former study and Smalltalk in the latter study. It is likely that some Smalltalk environment characteristics may have triggered the use of the function-centred strategy. Under this environment, methods performing the same function, e.g. initialisation, can be grouped by the programmer into a category of methods with the label « init », for example.

To conclude, we should note some thematic limitations of our review. We did not review work on OO education and learning environments even though there is some significant work in this area (see for example: Carroll, Rosson & Singley, 1993; Robertson, Carroll, Mack, Rosson & Alpert, 1993; Singley & Carroll, 1990) .

Topics which have not yet been addressed in the literature include OO program comprehension, on the one hand, and components retrieval, on the other hand. As regard OO program comprehension, we note that some studies are in progress (Burkhardt, Détienne & Wiedenbeck, 1997). Also, we regret that there are, as far as we know, almost no empirical studies on real novices. Future research could centre on these themes.

## Background

This paper is based on an invited talk presented by the author at ESP6 (Washington DC, January 5-7, 1996).

## Acknowledgements

We would like to express our thanks to Robert Rist for his helpful and constructive comments on a previous draft of this paper. Special thanks to Susan Wiedenbeck who has kindly accepted to improve the English of this paper.